\documentstyle[12pt]{article}
\def\by#1#2{{\displaystyle {#1}\over \displaystyle {#2}}}
\def\d{{\rm d}}

\begin{document}

\begin{flushright}
LC-TH-2001-019  \\
IMSc/2001/01/02 \\
IISc-CTS-01/01 \\
\today 
\end{flushright}

\begin{center}
{\Large \bf 
$J/\psi$ production through resolved photon processes at $e^+\,e^-$
colliders} \\ [0.3cm]
R.M. Godbole, \\
CTS, Indian Institute of Science, Bangalore 560 012, \\
D. Indumathi, \\
The Institute of Mathematical Sciences, CIT Campus, Chennai 600 113, \\
M. Kr\"amer \\
Department of Physics and Astronomy, University of Edinburgh, \\
Edinburgh, EH9 3JZ, Scotland \\ [1cm]
\end{center}

\noindent {\bf Abstract:} 

\noindent We consider $J/\psi$ photoproduction in $e^+\, e^-$ as well
as linear photon colliders. We find that the process is dominated by
the resolved photon channel. Both the once-resolved and twice-resolved
cross-sections are sensitive to (different combinations of) the colour
octet matrix elements. Hence, this may be a good testing ground for
colour octet contributions in NRQCD. On the other hand, the
once-resolved $J/\psi$ production cross-section, particularly in a
linear photon collider, is sensitive to the gluon content of the
photon. Hence these cross-sections can be used to determine the parton
distribution functions, especially the gluon distribution, in a photon,
if the colour octet matrix elements are known.

\section{Introduction}

There has been considerable interest in the production of $J/\psi$ at
various colliders ever since the large discrepancy between the measured
rate of $J/\psi$ production and the (much smaller) prediction of the
colour singlet (CS) model was first observed at the $p\, \overline{p}$
collider Tevatron \cite{TeV}. An analysis of the data \cite{ChoL} using
the NRQCD factorisation approach by Bodwin, Braaten and Lepage
\cite{BBL} yielded colour octet (CO) contributions which seemed almost
an order of magnitude larger than the CS term. However, later data
from the $e\, p$ collider HERA \cite{HERA} did not see the anticipated
excess, especially at large $z$ values (where $z$ is the inelasticity
variable). Analyses of both fusion \cite{CK} and fragmentation
\cite{KK} contributions to both direct and resolved photon
contributions to $J/\psi$ production at the HERA $e\, p$ collider have
been performed. In fact, the zero $p_T$ result has also been evaluated
to NLO \cite{Mangano}.

The measurement of the $J/\psi$ and $\psi(2S)$ polarisation at the
Tevatron \cite{TeVpol} also did not show the expected large
polarisation with increasing $p_T$ as predicted by NRQCD with a
dominant colour octet contribution \cite{Beneke,Braaten}.

This may be attributed to the larger uncertainty of the nonperturbative
colour octet matrix elements $\langle 0 \vert {\cal O}^{J/\psi}[n] \vert
0 \rangle$, that contribute in the large-$z$ region. More recently, it
has been proposed \cite{Haegler} that inclusion of the nonvanishing
transverse momenta of the colliding partons may drastically change the
cross-section, especially at large transverse momenta. In particular,
calculations in NRQCD are usually based on collinear factorisation. On
including the $k_\perp$ effects, it is possible to fit the octet matrix
element $\langle O^{J/\psi} [8,{}^3S_1] \rangle$ to a much smaller
value \cite{Haegler2} than that from NRQCD in the collinear limit.
This term contributed most substantially to the $J/\psi$ polarisation;
hence this may resolve the problem of the observed $J/\psi$ polarisation at
the Tevatron \cite{Haegler2}. However, the discrepancy between the CO
fits to the hadroproduction (Tevatron) and leptoproduction (HERA) data
remains. It is therefore interesting to estimate the CO as well as CS
contributions to $J/\psi$ production in various other processes.

Here, we examine the dependence of CO $J/\psi$ photoproduction on the
various NRQCD matrix elements in $e^+\,e^-$ and photon-photon
colliders. (Prompt production at $e^+\, e^-$ colliders has been
studied in Ref. \cite{prompt}). Apart from the direct contribution,
there are contributions from diagrams where either one or both of the
photons is resolved, so that the underlying parton structure is probed.
We are concerned here with these resolved photon contributions, the
direct contribution being small, as has already been observed for the
case of $\gamma\, \gamma$ colliders \cite{direct,JTkab}.

In particular, there are both colour singlet (CS) and colour octet
(CO) contributions to each of these processes. The CS cross section is
well known \cite{Baier}; in fact, it has long since been established
that the once resolved (1-res) photon contribution dominates the twice
resolved (2-res) photon contribution in the CS case; this was in fact
used to estimate the gluon content of the photon \cite{DreesG}. The
1-res case is similar to leptoproduction while the 2-res case is
analogous to hadroproduction in $p\, \overline{p}$ collisions; hence it
will be possible to examine both kinds of processes in a single
experiment. Also, effects of intrinsic $k_\perp$ should be different in
$\gamma\,\gamma$ scattering as compared to $e\,p$ or $p\,\overline{p}$
processes. In the context of the currently discussed $k_\perp$
factorisation as the solution to the observed $J/\psi$ polarisation at
the TeVatron, it would therefore be interesting to study $J/\psi$
production in these $\gamma\,\gamma$ processes.

These resolved processes have a very different topology from that of
the direct processes; hence they can be easily identified. For
instance, resolved photon processes have an extra (spectator) jet
occurring when a coloured parton of the photon interacts directly in
the hard scattering rather than the colour singlet photon itself.
Usually this jet is in the same direction as the parent photon (or
electron); indeed, it is analogous to the forward jet of remnants
produced from deep inelastic scattering off a hadron target. A
twice-resolved process, where both the photons are resolved into their
parton components will thus have two such jets. Hence direct, 1-res and
2-res processes can be separated event by event, based on the observed
topology.

The matrix elements in the CO case, with $n = [8,{}^3S_1],
[8,{}^1S_0]$, and $[8,{}^3P_J], J = 0,1,2$, are not as well established
as the CS ones and have been obtained from $J/\psi$ production at the
Tevatron \cite{ChoL,Sanchis,Leib}. Though these are estimated to be
about two orders of magnitude smaller than the CS matrix element, the
CO contributions are not expected to be small since they correspond to
diagrams of lower order in the strong coupling $\alpha_s$, or are
enhanced by $t$-channel gluon exchange, forbidden in the leading-order
colour singlet cross section.

We shall therefore compute the CS and CO contributions to the $J/\psi$
photoproduction cross-section at photon-photon colliders, using certain
reasonable estimates for the corresponding matrix elements. In the next
section, we will define the choice of kinematics and list the various
subprocesses that contribute to $J/\psi$ production at an $e^+\, e^-$
collider. Numerical results for the cross-section at LEP2 as well as a
future possible linear collider at $\sqrt{s} = 500$ GeV are presented
in Section 3. Section 4 discusses the contrasting ressults obtained for
a photon linear collider, where high intensity photon beams can be
obtained by scattering laser beams off electron beans. Numerical results
here are presented for the case $\sqrt s = 500$ GeV, along with some
discussions.

\section{$J/\psi$ photoproduction in $e^+\,e^-$ colliders}
\subsection{Kinematics and cross-sections}

$J/\psi$ can be produced via direct $\gamma\, \gamma$ interaction, or when
either or both of the photons are resolved into their partonic
constituents. We will refer to the direct interaction, and the once- and
twice- resolved photon processes as Direct, 1-res and 2-res processes
respectively. Both colour singlet (CS) and colour octet (CO)
subprocesses contribute to $J/\psi$ production in these three channels.
Also, $2 \to 2$ as well as $2 \to 1$ subprocesses contribute.
Specifically, they are

\noindent Direct:
\begin{eqnarray} \nonumber
\gamma \gamma & \rightarrow & (c \overline{c} ) \gamma \hbox{ \rm (CS)}~, \\
\gamma \gamma & \rightarrow & (c \overline{c} ) g \hbox{ \rm (CO)}~. \\
\nonumber
\end{eqnarray}
\noindent 1-res:
\begin{eqnarray} \nonumber
\gamma g_\gamma & \rightarrow & (c \overline{c} ) \hbox{ \rm (CO)}~, \\
\nonumber
\gamma g_\gamma & \rightarrow & (c \overline{c} ) g \hbox{ \rm (CS, CO)}~, \\
\gamma q_\gamma & \rightarrow & (c \overline{c} ) q \hbox{ \rm (CO)}~. \\
\nonumber
\end{eqnarray}
\noindent 2-res:
\begin{eqnarray} \nonumber
g_\gamma g_\gamma & \rightarrow & (c \overline{c} ) \hbox{ \rm (CO)}~, \\
\nonumber
q_\gamma \overline{q}_\gamma & \rightarrow & (c \overline{c} )
\hbox{ \rm(CO)}~, \\ \nonumber
g_\gamma g_\gamma & \rightarrow & (c \overline{c} ) g
\hbox{ \rm (CS, CO)}~, \\ \nonumber
g_\gamma q_\gamma & \rightarrow & (c \overline{c} ) g \hbox{ \rm (CO)}~, \\
q_\gamma \overline{q}_\gamma & \rightarrow & (c \overline{c} ) g
\hbox{ \rm(CO)}~.  \\ \nonumber
\end{eqnarray}
Note that the zero $p_T$ $2\to1$ contributions are purely CO. The 1-res
processes are analogous to those contributing to the $e\, p$ or $\gamma
\, p$ $J/\psi$ production processes at HERA, while the 2-res ones are
analogous to either the resolved $J/\psi$ photoproduction processes at
HERA or to $J/\psi$ production at the Tevatron. In both cases, the
parton densities in the proton are replaced by parton densities in the
photon for the case of interest. Hence modulo the difference in parton
densities, the production rate in the 1-res channel should reflect that
seen at HERA and the 2-res channel that at Tevatron. The $J/\psi$
production data from $e^+\,e^-$ collisions can therefore provide a
corroboration of the behaviour seen at $e\,p$ and $p\, \overline{p}$
colliders, and establish whether there is indeed a dominant CO
contribution in $J/\psi$ production at colliders.

The cross-section in the CM frame for the process $e^+\,e^- \to J/\psi \,
X$ is given by
$$
\by{\d^3 \sigma}{\d x_1 \d x_2 \d \hat{t}} = p_1(x_1) p_2(x_2) \by{\d
\hat\sigma}{\d \hat{t}} + (x_1 \leftrightarrow x_2)~,
$$
where 1 and 2 refer to the $e^+$ and $e^-$ respectively. Here $p_e(x)$
corresponds to $\gamma_e(x)$ for the case of the unresolved photon and
equals the convolution,
$$
p_e(x) = \int_x^1 \by{\d y}{y} \gamma_e(y) p_\gamma\left(x/y\right)~,
$$
in terms of the parton density $p_\gamma(x), p = q, g$, in the resolved
photon.
We use the Weiz\"acker-Williams approximation (WWa)
for the bremsstrahlung photon distribution from an electron:
\begin{equation}
f_{\gamma_e}(z) = \by{\alpha_{\rm em}}{2\pi} \left( \by{1+(1-z)^2}{z}
\log(q_{\rm max}^2/q_{\rm min}^2) + 2m_e^2 z \left(
\by{1}{q_{\rm max}^2} - \by{1}{q_{\rm min}^2}\right) \right)~,
\label{eq:WWa}
\end{equation}
where
$q_{\rm min}^2 = m_e^2 z^2/(1-z)$ and $q_{\rm max}^2 = (E\theta)^2 (1-z)
+ q_{\rm min}^2$. Here $z = E_\gamma/E_e$, $\theta$ is the angular
cut that ensures the photon is real, and $E = E_e = \sqrt{s}/2$. We use
a typical value of $\theta = 0.03$ in our analysis.
for $\sqrt{s} = 175$ GeV.

\section{Numerical results}
We recast the cross-section in terms of the hadronic variables,
$y_1$, $y_2$ and $p_T$ and compute the $p_T$ dependence of the
cross-section:
$$
\by{\d\sigma}{\d p_T}(e^+ e^- \to e^+ e^- J/\psi X)~.
$$
We use a common renormalisation and factorisation scale,
$q^2 = (m_c^2 + p_T^2)$ with $m_c = 1.5$ GeV and $\Lambda_{\rm QCD}^4
= 200$ MeV. We use the GRV leading order (LO) parametrisation \cite{GRV} for
the parton densities inside the photon. Similar results are obtained on
using the WHIT parametrisation \cite{WHIT} instead. 

We shall use the following reasonable choices for the
matrix elements which are consistent with the allowed values:
$\langle O^{J/\psi}[1,{}^3S_1] \rangle = 1.16$ GeV${}^3$,
$\langle O^{J/\psi}[8,{}^3S_1] \rangle = 10^{-2}$ GeV${}^3$,
$\langle O^{J/\psi}[8,{}^1S_0] \rangle = 10^{-2}$ GeV${}^3$,
$\langle O^{J/\psi}[8,{}^3P_0]\rangle /m_c^2 = 10^{-2}$ GeV${}^3$, where
the remaining $J$ values are fixed from symmetry:
$\langle O^{J/\psi}[8,{}^3P_J] \rangle = (2J+1)
\langle O^{J/\psi}[8,{}^3P_0] \rangle$. 

\begin{figure}[htb]
\centering
\vskip 8truecm

{\includegraphics{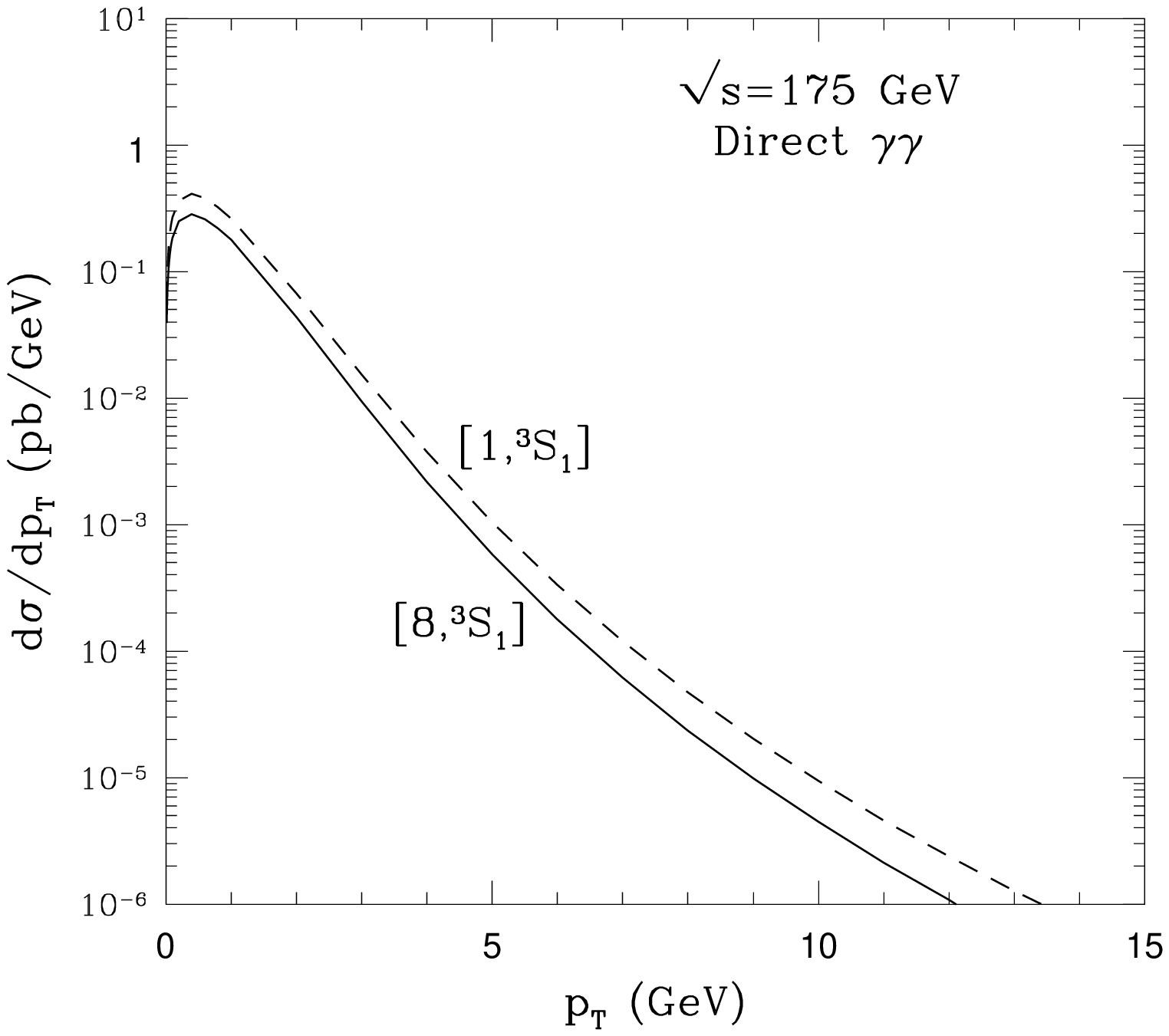}}

\caption{The direct $J/\psi$ photoproduction cross-section at LEP2 is
shown as a function of $p_T$. The CS and CO contributions are
separately shown.}
\label{fig:0res}
\end{figure}

We compute the $p_T$ dependence of the cross-section for the direct
$\gamma\,\gamma$, the 1-res photon and the 2-res photon cases. The CS
and CO cross-sections, $\d\hat{\sigma}/\d\hat{t}$, for all the
processes listed in eqs.~(1-3) are known \cite{direct,1-res,2-res}. The
results for the direct case are shown in Fig.~\ref{fig:0res}, where the
differential cross-section for the direct $\gamma\, \gamma$ interaction
\cite{direct} is plotted as a function of $p_T$. The $[8,{}^3S_1]$
octet matrix element that occurs here does not contribute dominantly to
this cross-section.

\begin{figure}[bht]
\centering
\vskip 8truecm

{\includegraphics{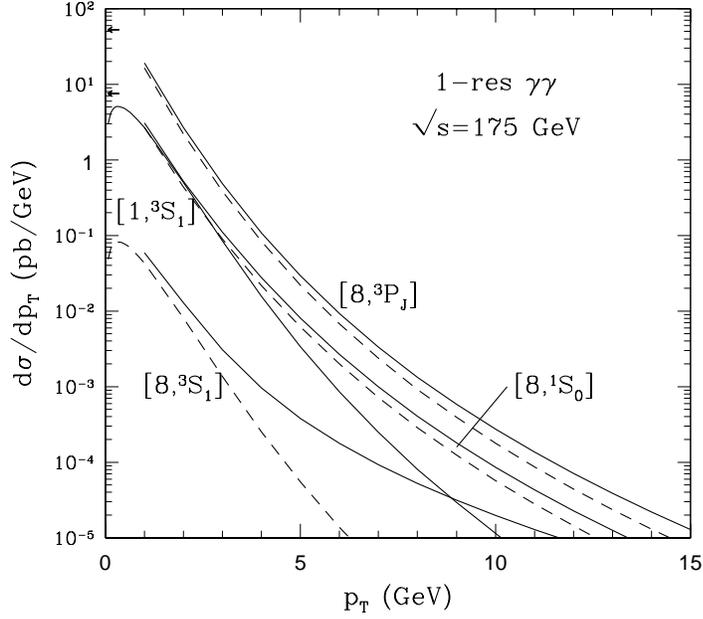}}

\caption{The $J/\psi$ photoproduction cross-section from once-resolved
(1-res) processes at LEP2 is shown as a function of $p_T$. The CS and
CO contributions are separately shown. The dashed (solid) lines
correspond to the gluon (total) CO cross-sections, the two differing
substantially only for the ${}^3S_1$ case. The arrows indicate the
zero $p_T$ $[8,{}^3P_J]$ and $[8,{}^1S_0]$ cross-sections (in pb),
arising from the corresponding $2\to 1$ subprocesses.}
\label{fig:1res}
\end{figure}

For the 1-res case, there are contributions from the ${}^3S_1$,
${}^1S_0$ and ${}^3P_J$ octet matrix elements apart from the singlet
${}^3S_1$ term \cite{1-res}. The $\gamma\,g$ interaction term is
expected to dominate this cross-section. These are shown as dashed
lines in Fig.~\ref{fig:1res}, where the individual contributions are
shown. The slope ($p_T$ dependence) of the octet $[8,{}^1S_0]$ and
$[8,{}^3P_J]$ terms is very similar with the ratio of the two
contributions ranging from about 0.15 near $p_T = 1$ GeV to about 0.32
near $p_T = 15$ GeV for $m_c = 1.5$ GeV. The slopes of the singlet and
octet ${}^3S_1$ terms are very different from these. Hence it may
be possible to separate the contribution involving the combination of
matrix elements $(\langle O^{J/\psi} [8,{}^1S_0 ] \rangle + 7 \langle
O^{J/\psi} [8,{}^3P_J ] \rangle/m_c^2 )$ from that of the ${}^3S_1$
terms, at small $p_T$. (At larger $p_T$, the cross-section drops off
rapidly).

A note about the cross-section as $p_T \to 0$. While the direct
cross-section remains finite for $p_T \to 0$, only the ${}^3S_1$
singlet and octet terms are finite for the 1-res case. The $2\to 2$
processes involving the ${}^1S_0$ and ${}^3P_J$ terms diverge in the
small-$p_T$ limit. However, precisely these processes have a finite CO
contribution from the $2\to 1$ zero $p_T$ processes; in fact, these
$2\to 2$ $\gamma g \to (c \overline{c})g$ processes at $p_T \to 0$ are
just these $2\to 1$ processes with a soft gluon emission. The apparent
divergence of the $2 \to 2$ cross-section at $p_T \to 0$ can be
resummed into a finite correction to the $2 \to 1$ lower order process
($K$-factor) \cite{Mangano}. Hence the $p_T = 0$ cross-section for the
$[8,{}^1S_0]$ and $[8,{}^3P_J]$ processes is within a $K$-factor of the
corresponding $2\to 1$ cross-section\footnote{The $2\to 1$ cross-sections
shown in Fig.~2 of Ref.~\cite{GIKold} should have been multiplied by
the corresponding matrix elements, that is, by a factor of $10^{-2}$.
Hence the conclusion drawn in that article about a substantial $2\to 1$
contribution at zero $p_T$ is wrong.} which is indicated by the arrows
marked in Fig.~\ref{fig:1res}.

Finally, the effect of including the $\gamma\, q$ terms is shown by the
solid lines in Fig.~\ref{fig:1res}. The quark contribution increases
with $p_T$ and is small. The one exception is the $[8,{}^3S_1]$
contribution which is significantly enhanced by inclusion of the quark
diagrams; however, this may still not be large enough to be observable.
The $\gamma\,q$ cross-section also diverges at small $p_T$. Here there
is no corresponding $2 \to 1$ lower order process. However, the
$J/\psi$ here is produced by fragmentation of a gluon; the soft
divergence at $p_T = 0$ must therefore be absorbed into the
fragmentation function in this case.

\begin{figure}[htb]
\centering
\vskip 8truecm

{\includegraphics{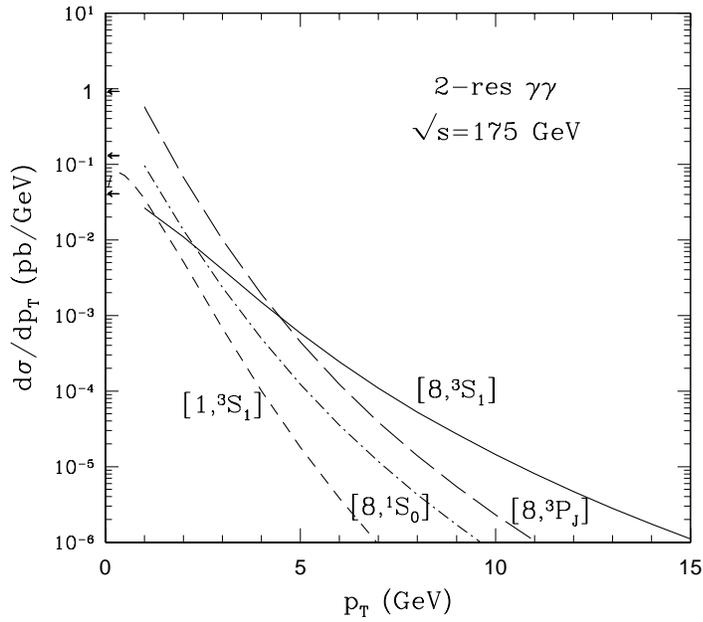}}

\caption{The $J/\psi$ photoproduction cross-section from twice-resolved
(2-res) processes at LEP2 is shown as a function of $p_T$. The CS and
CO contributions are separately shown. The arrows indicate the zero
$p_T$ $[8,{}^3P_J]$, $[8,{}^1S_0]$ and $[8,{}^3S_1]$ cross-sections (in
pb), arising from the corresponding $2\to 1$ subprocesses.}
\label{fig:2res}
\end{figure}

The subprocess cross-sections for 2-res processes are the same as those
for $p$-$\overline{p}$ collisions \cite{2-res} since the parton content
of both photons is resolved in this case. Contributions are from $gg$,
$gq$ and $q \overline{q}$ subprocesses. Here it turns out that the
octet $[8,{}^3S_1]$ term dominates at large $p_T$ as can be seen from
Fig.~\ref{fig:2res}. The $[8,{}^1S_0]$ and the $[8,{}^3P_J]$ terms
dominate at low $p_T$ and contribute in the same ratio as in the 1-res
case. Notice that the $( \langle O^{J/\psi}[8,{}^1S_0] \rangle + 7
\langle O^{J/\psi}[8,{}^3P_J]\rangle /m_c^2)$ contribution is much
larger than the CS term, unlike in the 1-res case. Hence, even if the
octet matrix elements are overestimated by a factor of 10, the CO
contribution is still substantial in the 2-res case. Note also that,
while the 2-res cross-section is only a few percent of the 1-res one,
it can be kinematically easily distinguished from the 1-res case and
can be analysed for its CO content. Hence it may be possible to
determine these CO matrix elements accurately through the 2-res
channel.  As in the 1-res case, the arrows in Fig.~\ref{fig:2res} at
$p_T = 0$ indicate the $2 \to 1$ contribution from the octet
$[8,{}^3S_1]$, $[8,{}^1S_0]$ and $[8,{}^3P_J]$ terms. The actual $p_T =
0$ cross-section will be within a $K$-factor of this (from the soft
limit of the corresponding $2 \to 2$ diagrams).  The CS term is finite
as $p_T \to 0$, as in the 1-res case. At a collider, it may be possible
to observe the zero $p_T$ $J/\psi$'s by reconstructing the leptonic
decay mode.

\begin{figure}[htb]
\centering
\vskip 8truecm

{\includegraphics{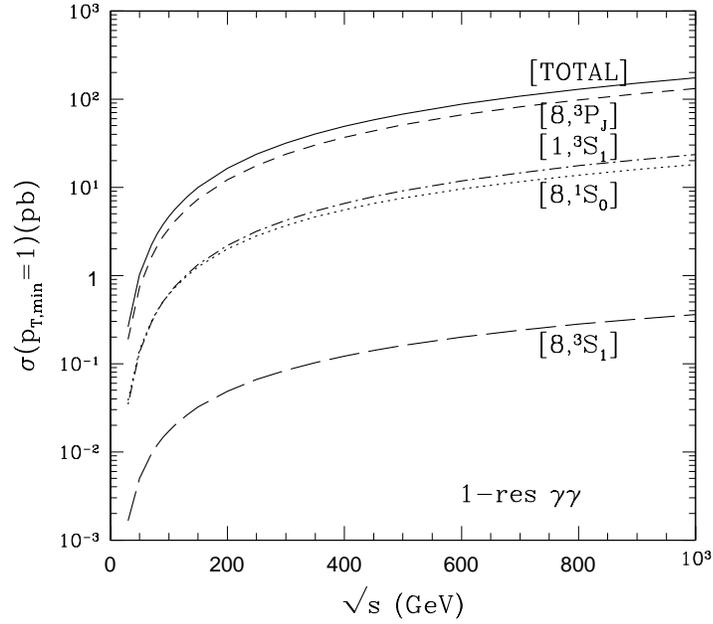}}

\caption{The 1-res $J/\psi$ photoproduction cross-section
integrated over $p_T$ from $p_{T,{\rm min}} = 1$ GeV, shown as a
function of $\sqrt{s}$. The CS, CO and total contributions are
separately shown.}
\label{fig:1restinteg}
\end{figure}

There exists substantial amount of data from LEP at $\sqrt s = 189$ GeV
as well; the results in this case are very similar to what is obtained
at the slightly smaller value of $\sqrt s$ used here. The variation of
the cross-section with the CM energy is shown in the next two figures.
The total 1-res cross-section (integrated from $p_{T,{\rm min}} = 1$
GeV to a kinematical maximum of $p_{T,{\rm max}} = \sqrt{s}/2$) is
shown in Fig.~\ref{fig:1restinteg} as a function of the centre of mass
energy, $\sqrt{s}$. The cross-section is small at lower energies, as
are available at colliders such as {\sc tristan} but increases
with $\sqrt{s}$. The cross-section at an $e^+\, e^-$ collider with
$\sqrt{s} = 500 $ GeV is $\sigma (P_{T,min} = 1 \hbox{ GeV}) = 68$ pb.
The number of $J/\psi$s seen will depend on the luminosity, and the
branching fraction of the cleanest decay mode, $J/\psi \to l^+ l^-$,
which is 6\%. However the ratio of the octet $[8,{}^1S_0]$ to
$[8,{}^3P_J]$ terms is a fairly steady 0.15 over a large range of
$\sqrt{s}$.  Hence it is possible that a combination of 1-res and 2-res
processes at $e^+\,e^-$ colliders can help determine the universal CO
matrix elements occurring in $J/\psi$ production.

\begin{figure}[htb]
\centering
\vskip 8truecm

{\includegraphics{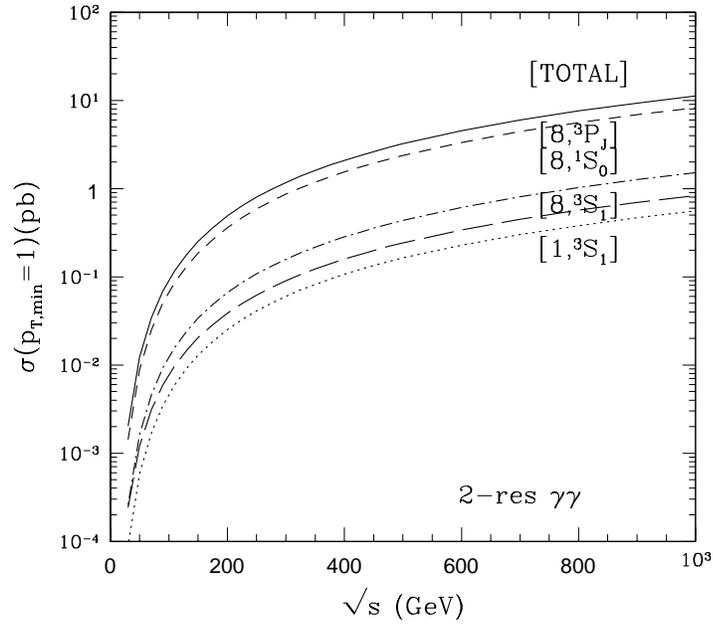}}

\caption{As in Fig.~\ref{fig:1restinteg} for the integrated 1-res
cross-section, but for the 2-res case.}
\label{fig:2restinteg}
\end{figure}

The corresponding total cross-section for the 2-res case is plotted in
Fig.~\ref{fig:2restinteg}. In general, the inclusion of CO terms does
not affect the result that the 1-res dominates the 2-res contributions.
Also, we find that the CO contribution is much larger than the CS one;
this may also reflect the fact that we have used octet matrix elements
from the Tevatron fits which may overestimate $J/\psi$ production at
HERA.  However, independently of this, the 1-res contribution
dominates. This is in contrast to the $e$-$p$ case, for example, at
HERA, where the resolved photon contribution (corresponding to the
2-res term in $\gamma\, \gamma$ collisions) is an appreciable fraction of
the direct one (corresponding to the 1-res term of $\gamma\, \gamma$
collisions) \cite{CK2,GRS}.

Realistic acceptance cuts on the lepton angle and $p_T$ should reduce
the event rates at {\sc tristan} by approximately a factor two but only by
about 10\% in the case of LEP2. Accurate estimates will be presented in
a future work.

In the case of larger $p_T$ events, the situation is not so promising,
since the production rate falls very rapidly with $p_T$. What may be
interesting to examine is whether rapidity cuts will enhance the colour
octet contribution or else distinguish in some way the CO from the CS
part. We leave this question to future work.

Finally, we remark that there is a further uncertainty in $e^+\, e^-$
collisions compared to $e$-$p$ collisions since the parton densities in
the photon are not as well known as those in the proton.

\begin{figure}[bht]
\centering
\vskip 8truecm

{\includegraphics{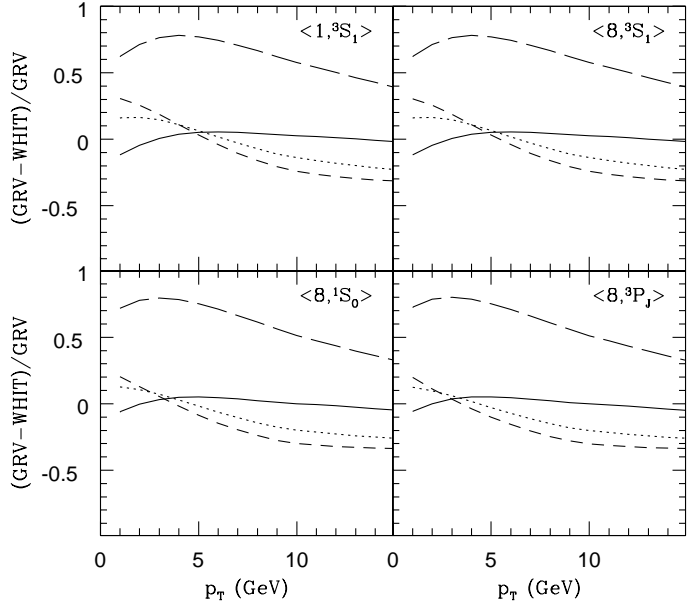}}

\caption{The variation of the $J/\psi$ photoproduction cross-section from once-resolved
(1-res) processes at LEP2 for different parametrisations of the photon
density is shown as a function of $p_T$. 
The solid, dotted, dashed and long-dashed lines
correspond to the WHIT1,2,3,4 parametrisations for the gluon density.}
\label{fig:1res_par}
\end{figure}

The dependence of the cross-section on the choice of parametrisation is
shown in Fig.~\ref{fig:1res_par}. The four panels show the sensitivity
of the individual gluon contribution only for the different 1-res
singlet and CO contributions when the WHIT rather than the GRV
parametrisations are used. The WHIT1 gluon is closest to the GRV gluon.
The WHIT2,3 are smaller at $x > 0.1$ while the WHIT4 has a
gluon that is twice that of WHIT1. While the corrections are rather large,
especially for the WHIT4 density, where it exceeds 50\%, the $p_T$
dependence is the same (in all 4 panels) for a given parametrisation
for all the CS and CO terms and is rather flat. Unless the CS and CO
matrix elements are known to precision, therefore, it may not be
possible to distinguish the different parametrisations from the 1-res
cross-section.

\begin{figure}[htb]
\centering
\vskip 8truecm

{\includegraphics{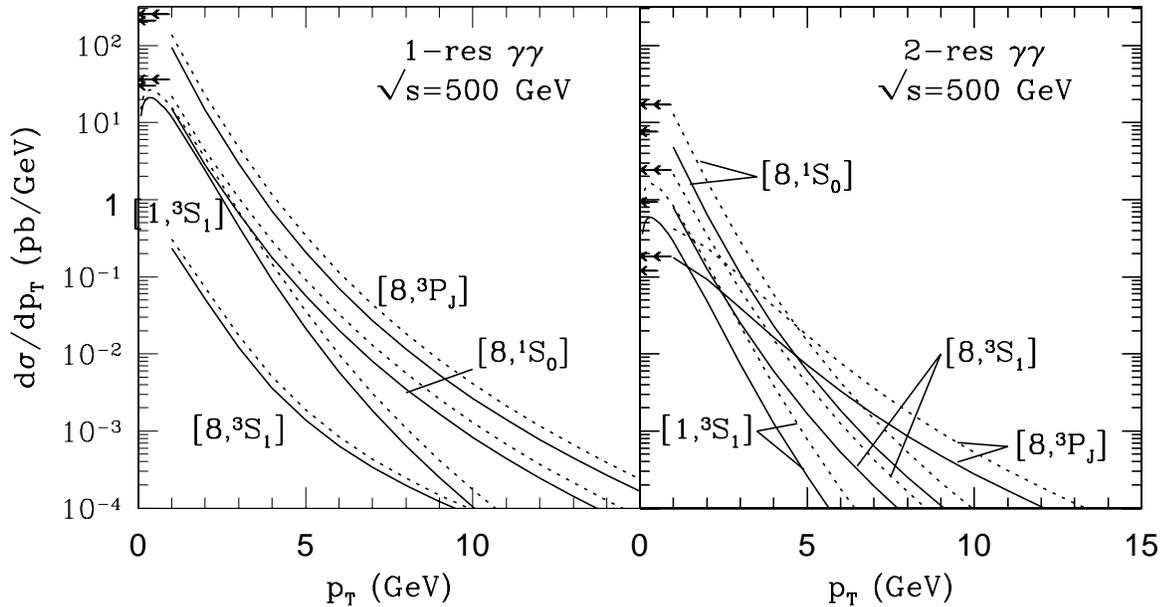}}

\caption{The total CS and CO 1-res and 2-res $J/\psi$ photoproduction
cross-sections shown as a function of $p_T$ for a future $e^+\,e^-$
linear collider at $\sqrt{s} = 500$ GeV. Solid and dotted lines
(correspond to the use of GRV and WHIT4 parametrisations for the parton
densities in a photon. The zero $p_T$ $2 \to 1$ contributions (in pb)
are indicated by (double) arrows for the (WHIT) GRV cases respectively.}
\label{fig:res_500}
\end{figure}

This can be seen from Fig.~\ref{fig:res_500} where the 1-res and 2-res
cross-sections are shown for a future linear collider at $\sqrt s =
500$ GeV. The total cross-section is about an order of magnitude larger
than at LEP2; however, the other features (such as the $p_T$ dependence
of the various CS and CO contributions) remain the same when we go to
larger $\sqrt s$ values. The sensitivity of the cross-section to the
choice of parton densities in a photon is also shown in this figure.
There is not much difference between the predictions from the GRV
\cite{GRV} and WHIT4 \cite{WHIT} parton distribution sets for the 1-res
case. However, since both photons are resolved into their partonic
content in the 2-res case, the predictions are more sensitive to the
densities in the 2-res case. It is seen that the cross-sections are
systematically higher when the WHIT4 parametrisation is used than with
the GRV set. However, the shape ($p_T$ dependence) remains roughly the same,
independent of choice of parametrisation.

\section{$J/\psi$ production from a photon linear collider}
High intensity photon beams can be obtained by back-scattering of laser
beams off electron beams. Such a photon linear collider can have high
energies of $\sqrt{s} = 500$--1000 GeV and very high luminosity. Hence
there has recently been a great deal of interest in such colliders. 

The $J/\psi$ production processes here are the same as in $e^+\,e^-$
colliders. Since the photons are accelerated by back-scattering, they
are distributed very differently from the WWa case. In place of
eq.~(\ref{eq:WWa}) for the WWa photons, we have
\begin{equation}
\gamma_{\rm laser}(z) = \left(\by{1}{1-z} + 1 - z - 4r (1-r)\right)
\by{1}{\sigma_c}~,
\label{eq:laser}
\end{equation}
where $r = z/(\kappa (1-z))$ and the maximum energy of the photon is
limited to $z_{\rm max} = \kappa/(1+\kappa)$, where the
dimensionless variable, $\kappa$, is given by,
$$
\kappa = \by{4E_b E_0}{m_e^2} \cos\theta/2~,
$$
for an electron beam of energy $E_b$, a laser of energy $E_0$ and
$\theta$ the angle between them. Here,
$$
\sigma_c = \log(1+\kappa) + z_{\rm max}^2 \left( \by{\kappa+2}{2\kappa}
\right) + \by{4}{\kappa} \left(z_{\rm max} + \kappa - 2 \log(1+\kappa)
\right)~,
$$
and we choose $\kappa = 4.83$ to avoid background from pair creation
processes, $\gamma\gamma \to e^+ e^-$, in the collision.

\begin{figure}[htb]
\centering
\vskip 8truecm

{\includegraphics{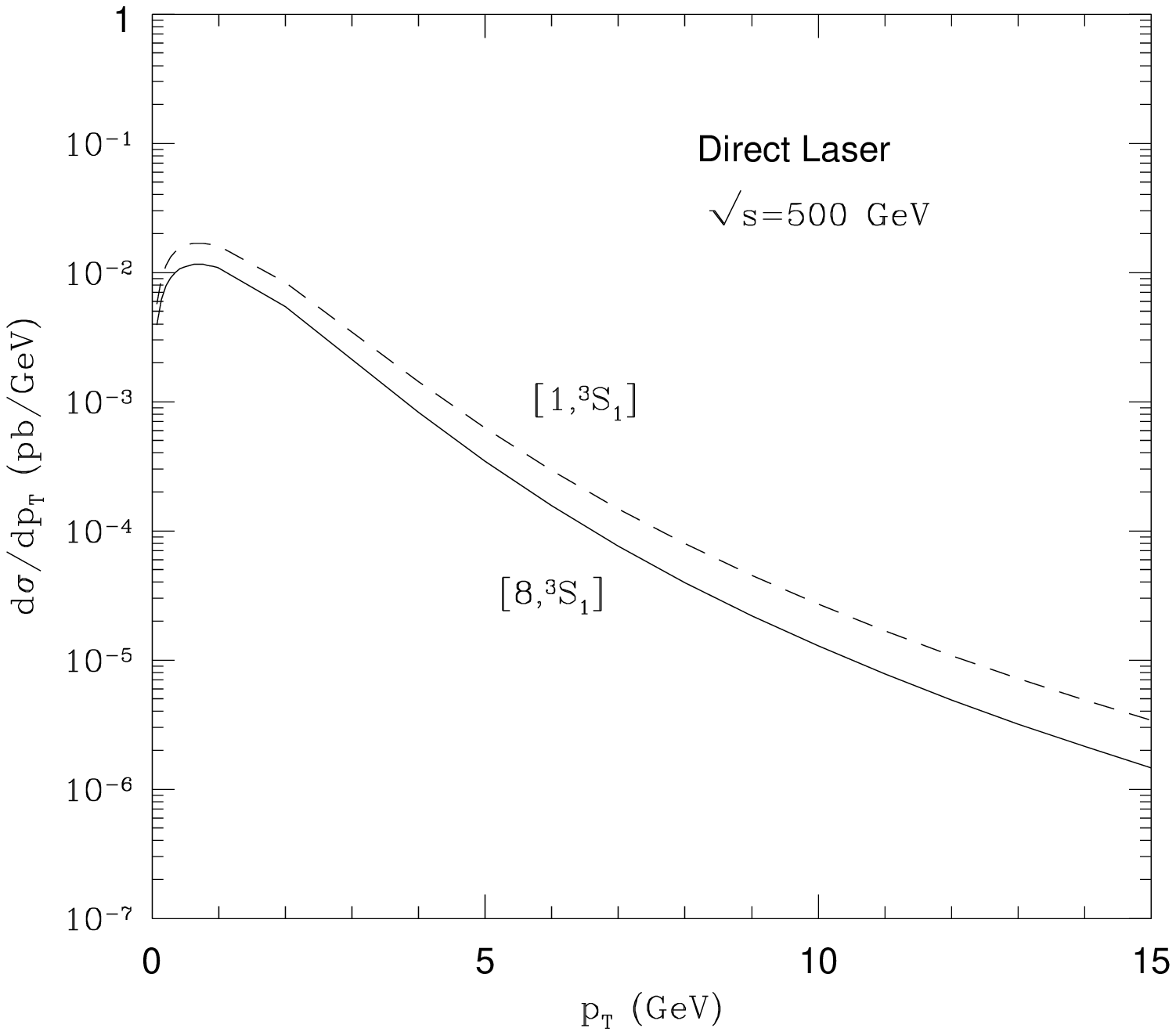}}

\caption{The same as Fig.~\ref{fig:0res}, but for a laser
backscattered photon at $\sqrt{s} = 500$ GeV.}
\label{fig:0reslaser}
\end{figure}

\begin{figure}[htb]
\centering
\vskip 8truecm

{\includegraphics{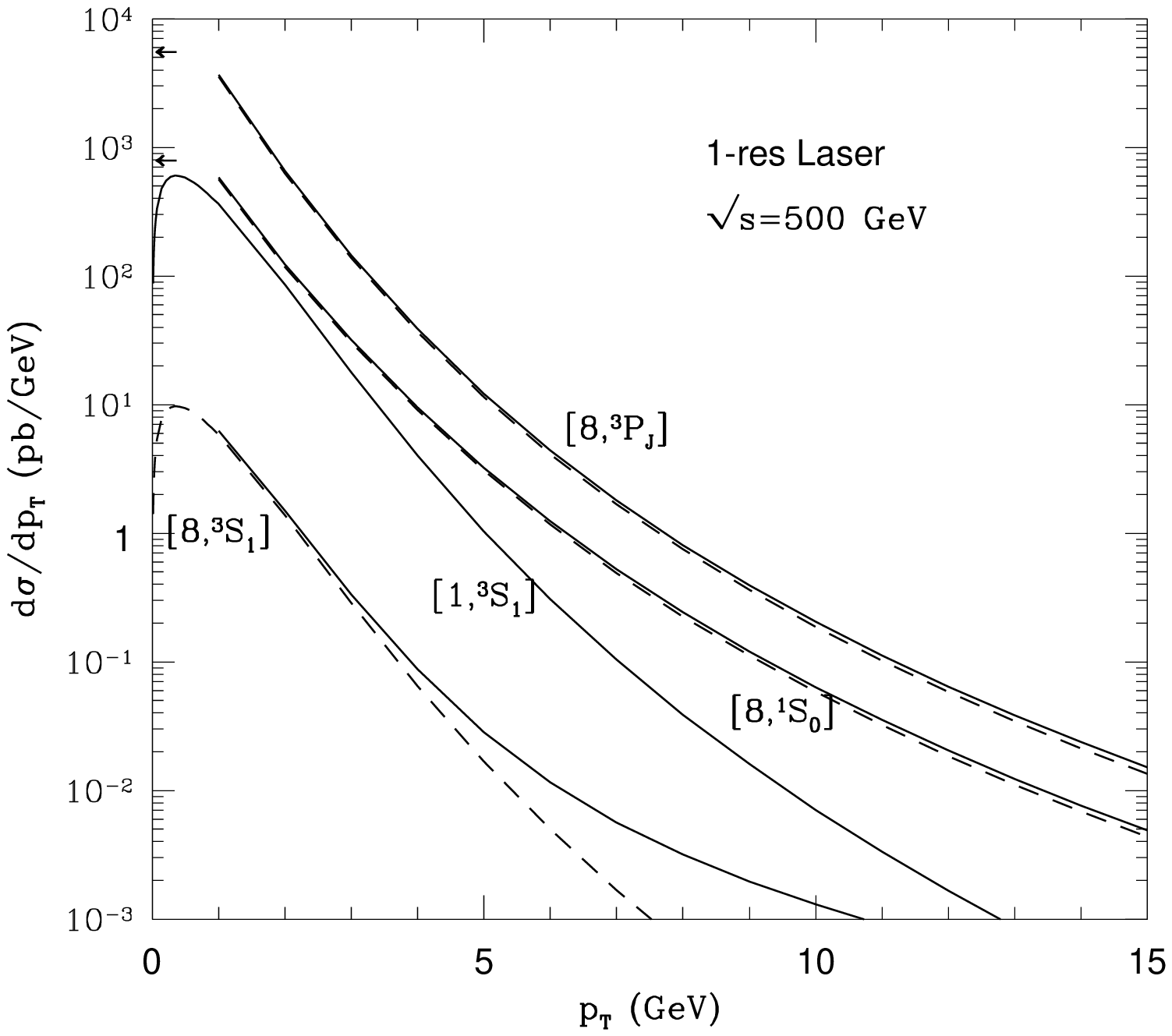}}

\caption{The same as Fig.~\ref{fig:1res}, but for a laser
backscattered photon at $\sqrt{s} = 500$ GeV.}
\label{fig:1reslaser}
\end{figure}

\begin{figure}[htb]
\centering
\vskip 8truecm

{\includegraphics{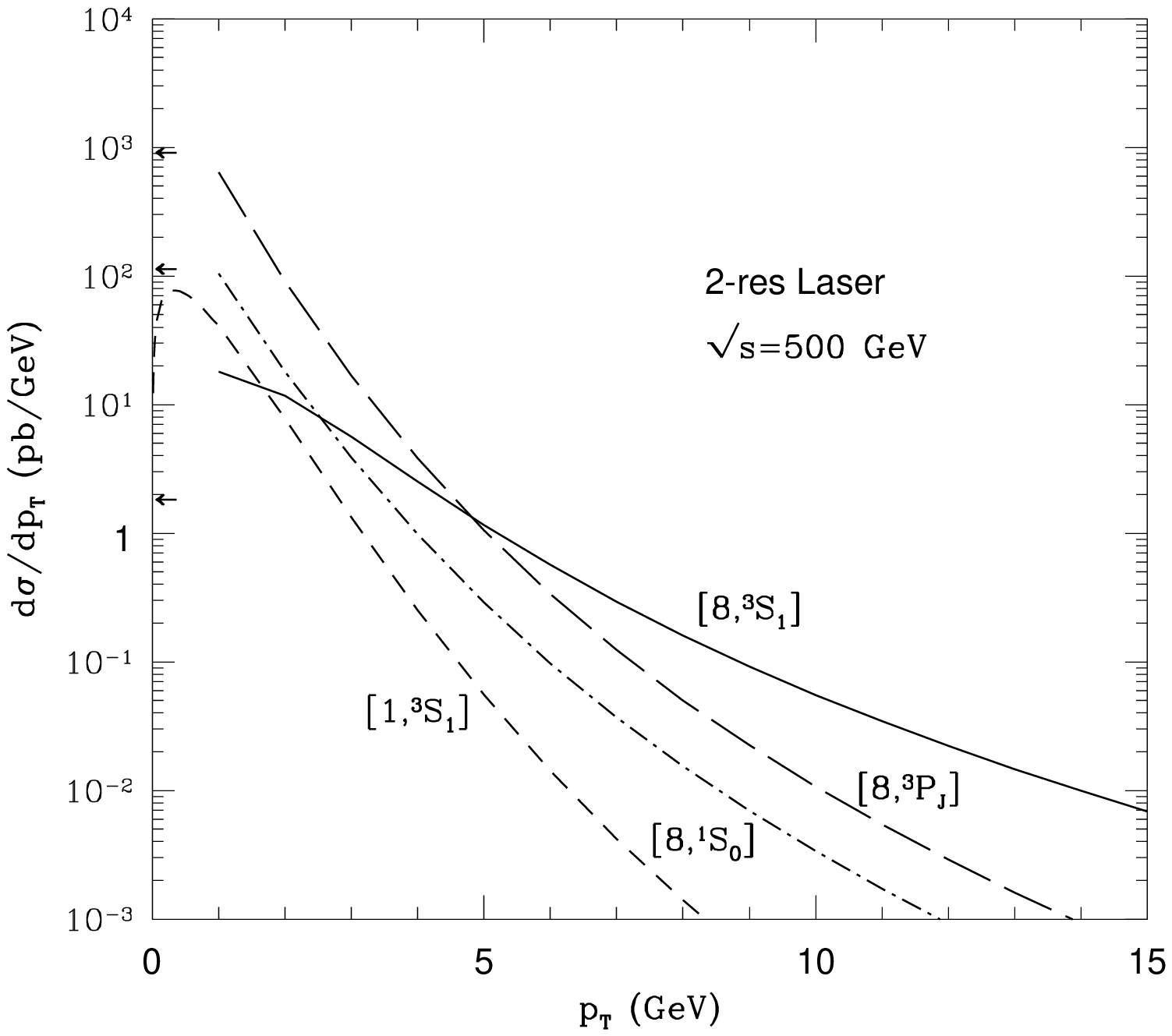}}

\caption{The same as Fig.~\ref{fig:2res}, but for a laser
backscattered photon at $\sqrt{s} = 500$ GeV.}
\label{fig:2reslaser}
\end{figure}

We again use the GRV parametrisation \cite{GRV} for parton distributions
in the photon and compute the same cross-section, but for the laser
back-scattered photon-photon scattering. That is, the subprocesses are
the same as for the $e^+e^-$ case, but the laser photon distribution
given in eq.~(\ref{eq:laser}) is to be used instead of the WWa
distribution. We present the results for such a future collider with
$\sqrt{s} = 500$ GeV in Figs.~\ref{fig:0reslaser}, \ref{fig:1reslaser},
\ref{fig:2reslaser}. Since the subprocesses are the same as in the
$e^+e^-$ case, the $p_T$ dependences are the same as before, with the
same behaviour of the octet $[8,{}^1S_0]$ and $[8,{}^3P_J]$ terms. The
advantage here is in the event rate which is much larger than in
$e^+e^-$ colliders, as can be seen from the much larger cross-section
in this case.  Furthermore, the direct contribution in photon colliders
is much smaller (by about two orders of magnitude) than in $e^+e^-$
colliders. Hence $J/\psi$ production at photon colliders will be
dominated by the resolved contributions. Photon colliders will
therefore be good sites for testing the colour octet contribution and
obtaining the octet matrix elements that occur in $J/\psi$ production.
Furthermore, the quark contribution to the 1-res case is negligible
here. Hence the 1-res cross-section is proportional to the gluon content
of the photon.

We have ignored the contribution to the cross-section from $\chi$
feed-down; however, with sufficient data, it may be possible to
separate the prompt $J/\psi$ production rate from these decay modes. It
may still be hard to separate out the individual octet $[8,{}^1S_0]$
and $[8,{}^3P_J]$ contributions in these processes.

In conclusion, $J/\psi$ photoproduction at both $e^+\, e^-$ as well as
photon linear colliders can prove to be a sensitive testing ground to
determine the colour octet contribution in $J/\psi$ production. This,
in comparison with the data from $p\, \overline{p}$ and $e\, p$
colliders, can help determine the colour octet matrix elements involved
in $J/\psi$ production. It is also possible to use the shape of the
$p_T$ spectrum to determine the various contributions. Turning the
problem around, if the NRQCD matrix elements for the process are
determined by other experiments, it is possible to use the measured
$J/\psi$ photoproduction cross-sections as proposed in this paper, to
determine the parton distribution functions in a resolved photon.

\paragraph{Acknowledgement}: MK thanks M. Cacciari, B.A. Kniehl and F.
Maltoni for useful discussions. The work of RMG and DI was supported in
part under CSIR grant No. 3 (745) 94-EMR-II. The authors also wish to
thank the DST and the organisers of the fifth Workshop on High Energy
Phenomenology (WHEPP-5), where these results were first presented.

\vspace{10cm}

\end{document}